\documentclass[structabstract]{aa}  
\usepackage{graphicx}
\def\cs{{c_{\rm s}}}
\newcommand{\dfrac}[2]{\frac{\displaystyle{#1}}{\displaystyle{#2}}}
\newcommand{\stck}[1]{\stackrel{#1}{\longrightarrow}}

\begin{document}
   \title{Effects of accretion flow on the chemical structure in the
inner regions of protoplanetary disks} 

%   \subtitle{I. Overviewing the $\kappa$-mechanism}

   \author{H. Nomura\inst{1}
\thanks{Now at Department of Astronomy, Graduate School of Science,
Kyoto University, Kyoto 606-8502, Japan},
 Y. Aikawa\inst{2}, Y. Nakagawa\inst{2}
          \and
          T.J. Millar\inst{1}
%          C. Ptolemy\inst{2}\fnmsep\thanks{Just to show the usage
%          of the elements in the author field}
          }

   \institute{Astrophysics Research Centre, School of Mathematics \&
Physics, Queen's University Belfast, Belfast BT7 1NN, UK\\
              \email{h.nomura@qub.ac.uk}
         \and
Department of Earth and Planetary Sciences, 
Kobe University, 1-1 Rokkodai-cho, Nada, Kobe 657-8501, Japan
%             University of Alexandria, Department of Geography, ...\\
%             \email{c.ptolemy@hipparch.uheaven.space}
%             \thanks{The university of heaven temporarily does not
%                     accept e-mails}
             }

%   \date{Received September 15, 1996; accepted March 16, 1997}

 \abstract
{}
{We have studied the dependence of the profiles of molecular abundances
and line emission on the accretion flow in the hot ($\ga 100$K)
inner region of protoplanetary disks.} 
{The gas-phase reactions initiated by evaporation of the ice mantle on
dust grains are calculated along the accretion flow. 
We focus on methanol, a molecule that is formed predominantly through
the evaporation of warm ice mantles, to show how the abundance profile
and line emission depend on the accretion flow.} 
%and the line emission of methanol is simulated using the obtained
%abundance profile.
{Our results show that some evaporated molecules keep high
abundances only when the accretion
velocity is large enough, and that methanol could be useful as
a diagnostic of the accretion flow  by means
of ALMA observations at the disk radius of $\la 10$AU.}
{} 


 \authorrunning{H.Nomura et al.}
 \titlerunning{Effects of accretion flow on inner disk chemistry}
   \maketitle
%
%________________________________________________________________

\section{Introduction}

Recent observations have detected a variety of
molecular lines towards disks around T Tauri stars
%(e.g., \cite{dut97}; \cite{thi04}; \cite{qi08})
%\citep[e.g.,][]{dut97}
%\citep[e.g.,][]{dut97,thi04,qi08}. 
(e.g., Dutrey et al. 1997; Thi et al. 2004; Qi et al. 2008). 
Existing millimetre/sub-millimetre 
observations, which have relatively low spatial resolution, can trace
only the outer 
region ($\ga 50$AU) of the disk, while near-infrared observations with high
sensitivity or high spectral resolution probe molecular lines from 
the planet-forming region in the disks (Lahuis et
al. 2006; Gibb et al. 2007; Carr \& Najita 2008). 
The forthcoming Atacama Large Millimeter/sub-millimeter Array (ALMA),
with both high sensitivity and high spatial resolution, will make
it possible to observe various molecular lines from the inner
regions of the disks. 
%(e.g., Semenov et al. 2008).

Although many models on the chemical structure of young circumstellar
disks focus on the outer disk (e.g., Aikawa et al. 2002; Willacy 2007),
%Vasyunin et al. 2008 
the chemistry in the inner disks has also been studied (e.g., Markwick
et al. 2002). 
In cold pre-stellar cores, observations suggest that many gas-phase molecules,
including CO, are frozen onto dust grains (e.g.,
Caselli et al. 1999). Young stars and circumstellar disks are thought to
be formed as a result of the collapse of such molecular cloud cores.
In addition, observations of molecular lines towards disks suggest that the
gas-phase molecules are frozen onto grains near the midplane of cold
outer disks (e.g., Dutrey et al. 1997). However, 
in the inner region of the disk the dust temperature is high due to the
irradiation from the central star, and the ice mantles 
%molecules 
are expected to evaporate into the gas
by analogy with the molecules observed towards star-forming cores,
so-called hot molecular cores and hot corinos
(e.g., Millar 1993; Ceccarelli et al. 2007),
%hot cores and hot corinos (high and low mass sfarforming cores), 
%which are thought to be formed as a result of the icy mantle
%evaporation and the subsequent gas-phase reactions 
so that the chemistry in the inner
disk will be characterized by this ice mantle evaporation.

%It is likely that the gas and dust in young
%circumstellar disks are accreting towards a central star and supplying
%an additional mass to the star. 
Accretion flow towards a central star is
important for planet formation in the disk because it is one
of the possible mechanisms for the dispersal of the gas, 
(e.g., Hollenbach et al. 2000), 
while the amount
of the gas in the disk influences the coagulation and settling of dust 
grains, which lead to planetesimal formation, and
also controls gaseous planet formation as well as the orbital motion
of the planets 
(e.g., Nakagawa et al. 1986; 
%Hayashi et al. 1985; 
Kominami \& Ida 2002; Papaloizou et al. 2007). 
Meanwhile theory has suggested that the planet forming region
in the disks could be magnetorotationally stable, that is, there could be
no driving source of the accretion flow (e.g., Sano et al. 2000).
Therefore, it will be useful if we can identify observational 
evidence for the accretion flow in the inner disk in order to
understand the planet formation processes.

If the gas and dust in the disks are accreting towards the central star, 
the ice mantle molecules, which are frozen onto dust grains in the cold
outer disk, will evaporate when they are transported into the hot inner
disk.  
The evaporated molecules will be destroyed by chemical reactions
while they are migrating inwards along the accretion flow.
Thus, the distributions of the molecules will trace the accretion
velocity, depending on whether the timescale of the chemical
reactions is shorter or longer than the accretion time. Hence, it
will be possible to check the existence of accretion flow
%the existence of accretion flow 
from observations of the molecular distributions in the inner disk.
%comparing the timescale of the chemical reactions and the accretion time,

In this work we study the effect of the accretion flow on the chemical
structure 
%and the observational properties 
of the inner region of 
young disks by modelling the chemistry, initiated by the ice evaporation, along
the accretion flow. 
And we examine the effect on the observational properties,
especially focusing on 
a transition line of methanol, which is an abundant ice molecule in many  
high and low mass star forming cores (e.g., Gibb et al. 2004; Boogert et
al. 2008). Methanol, which is difficult to form efficiently in the gas
phase, is also known to be more abundant in the gas-phase around high
and low mass young stars as well as comets (e.g., Macdonald et
al. 1996; Sch\"oier et al. 2002; Bockelee-Morvan et al. 1991) 
by more than orders of magnitude compared with cold dark clouds 
(e.g., Ohishi et al. 1992). 
In addition, laboratory experiments have shown that methanol can 
be formed in ice mantle with reasonably short timescale 
as a result of hydrogenation of carbon monoxide 
%and the observed abundances of icy methanol can be
%reproduced by the reaction with a reasonable timescale 
(e.g., Watanabe et al. 2006). 
Therefore, methanol is thought one of the most probable components of
grain mantle molecules.

In the following section we introduce the physical and
chemical models of the disk. We present the resulting molecular
abundance profiles of the inner disk 
%and a brief comparison between our results and the recent infrared observations 
in Sect.~3 and the line
emission of methanol in Sect.~4, using the models with different
accretion velocities. Influence of some assumptions on the results are
discussed in Sect.~5, and
%Finally, 
the results are summarized in Sect.~6.

%__________________________________________________________________

\section{Models}

We model an axisymmetric disk surrounding a central star with a mass of
$M_*=1.5M_{\odot}$, a radius of $R_*=2R_{\odot}$, 
and a temperature of $T_*=6000$K, whose luminosity ($\sim 5L_\odot$)
is relatively higher than the averaged luminosity of typical T Tauri
stars ($\sim 1L_\odot$) (e.g., D'Alessio et al. 2005).
It is observationally known that many T Tauri stars emit strong X-ray
and UV radiation.
We adopt a model with black body ($T_*=6000$K) plus thermal
bremsstrahlung ($T_{\rm br}=2.5\times 10^4$K) emission for UV radiation
($L_{\rm FUV}= 7\times 10^{31}$ergs s$^{-1}$; see Nomura \& Millar 2005,
hereafter NM05) and a simple thermal bremsstrahlung 
emission model with $T_{\rm X}=1$keV and $L_{\rm X}=10^{30}$ ergs
s$^{-1}$ for X-rays (Glassgold et al. 2004; Garmire et al. 2000).

\begin{figure}
 \centering
 \includegraphics[width=8cm]{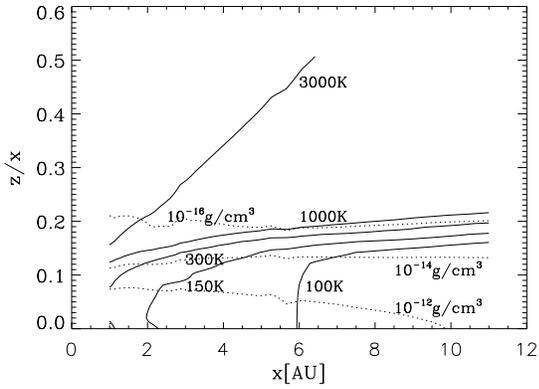}
 \caption{Contour plots of the gas temperature ({\it solid lines}) and
 density ({\it dotted lines}) distributions in the $z/x$ vs. $x$ plane
 for the fiducial model with $\dot{M}=1.0\times 10^{-8} M_{\odot}$
 yr$^{-1}$.} 
 \label{f0}
\end{figure}

The temperature and density distributions 
of gas and dust
in the disk are obtained
self-consistently as described in previous papers; see 
%Nomura \& Millar 2005, hereafter
NM05 and Nomura et al. (2007) for details.
Dust temperature is obtained by the iterative radiative transfer
calculation (see Nomura 2002),
where an initial dust temperature profile is calculated using
the variable Eddington factor method (Dullemond et al. 2002) 
in this work.
%We adopt a model with black body ($T_*=6000$K) plus thermal
%bremsstrahlung ($T_{\rm br}=2.5\times 10^4$K) emission for UV radiation
%($L_{\rm FUV}= 7\times 10^{31}$ergs
%s$^{-1}$; see NM05) and a simple thermal bremsstrahlung
%emission model with $T_{\rm X}=1$keV and $L_{\rm X}=10^{30}$ ergs
%s$^{-1}$ for X-rays (Glassgold et al. 2004; Garmire et al. 2000).
%The dust temperature is obtained by assuming local radiative equilibrium
%with the axisymmetric two-dimensional radiative transfer calculations
%(Nomura 2002). The variable Eddington factor method (Dullemond et
%al. 2002) is adopted for obtaining an initial dust temperature profile
%for the iterative radiative transfer calculations in this work.
The disk is assumed to accrete towards the central star with a constant
mass accretion rate, and the $\alpha$-viscous model with $\alpha=0.01$ is
adopted. The dust model which reproduces the observational extinction
curve of dense clouds (see NM05) is adopted here. 
The resulting gas temperature and density profiles for a fiducial model
with the mass accretion rate of $\dot{M}=1.0\times 10^{-8}M_{\odot}$
yr$^{-1}$ are plotted in Figure~\ref{f0}. 
%The line around $x=1$AU and
%$z/x=0.4-0.5$ in the figure shows the disk surface in this model.

%__________________________________________________ Table 1
\begin{table}
\caption[]{Initial fractional abundances with respect to total
hydrogen nuclei.}\label{T1}
$$ \begin{array}{p{0.11\linewidth}c|p{0.12\linewidth}c} \hline 
 Species & {\rm Abundance} & Species & {\rm Abundance} \\ \hline
 H$^+$ & 1.0\times 10^{-11} & CO & 1.3\times 10^{-4} \\
 He$^+$ & 2.5\times 10^{-12} & CO$_2$ & 3.0\times 10^{-6} \\
 H$_3^+$ & 1.0\times 10^{-9} & H$_2$CO & 2.0\times 10^{-6} \\
 Fe$^+$ & 2.4\times 10^{-8} & CH$_3$OH & 2.0\times 10^{-7} \\
 He & 1.0\times 10^{-1} & C$_2$H$_5$OH & 5.0\times 10^{-9} \\
 S & 5.0\times 10^{-9} & O$_2$ & 1.0\times 10^{-6} \\
 Si & 3.6\times 10^{-8} & H$_2$O & 2.8\times 10^{-4} \\
 C$_2$H$_2$ & 5.0\times 10^{-7} & N$_2$ & 3.7\times 10^{-5} \\
 CH$_4$ & 2.0\times 10^{-7} & NH$_3$ & 6.0\times 10^{-7} \\
 C$_2$H$_4$ & 5.0\times 10^{-9} & H$_2$S & 1.0\times 10^{-7} \\
 C$_2$H$_6$ & 5.0\times 10^{-9} & OCS & 5.0\times 10^{-8} \\ \hline
\end{array}
$$ 
\end{table}

For the chemical model we calculate the time-dependent gas-phase
reactions along stream lines of steady accretion flow as 
\begin{equation}
\dfrac{\partial (n_iv_{\rm acc})}{\partial l}=\sum_{j}k_{ij}n_{j}+\sum_{j,k}k_{ijk}n_{j}n_{k},
\end{equation}
where $n_{i}$ is the number density of the species, $i$, and 
the right hand side of the equation describes the formation and
destruction of the species, $i$, due to cosmic-ray/photo-chemistry (the
first term) and two-body reactions (the second term). Three-body
reactions are not included, since 
they will not affect the result very much in the gas temperature range
we treat in this work (e.g., Willacy et al. 1998).
We assume that the stream lines, $l$, are simply described as
$z=sH$, where $H=\cs_0/\Omega_{\rm K}$ is the disk scale height, and
$\cs_0$ the sound speed at the disk midplane, $\Omega_{\rm K}$ the
Keplerian frequency. We set 60 grids for $0\leq s\leq 3$. 
%in the calculations in Sect.~4.
%The total mass conservation is applied in order to solve the
%equations, 
The equations are solved together with the continuity
equation along the stream lines. 
The accretion velocity, $v_{\rm acc}$, is simply given by
$v_{\rm acc}=\dot{M}/(2\pi\Sigma x)$, where $\dot{M}$ is the mass
accretion rate, $\Sigma$ the surface 
density of the disk, and $x$ the distance from the central star.
Turbulent mixing is not taken into account in this paper (see
discussions in Sect.~5).

The chemical network consists of 208 species and is connected by 2830
reactions, in which the reaction rate coefficients are taken from the
UMIST RATE06 database (http://www.udfa.net/)
(Woodall et al. 2007). The X-ray ionisation is
simply modelled by analogy with the cosmic-ray ionisation (see e.g., Aikawa
et al. 1999 for a more detailed model). 
X-rays will ionise the gas and induce photoreactions, similar to
cosmic-rays; 
we adopt the rates enhanced by $\zeta_X/\zeta_{\rm CR}$ for the
reactions, where the X-ray ionisation rate, $\zeta_X$, is calculated
using the X-ray flux at each position in the disk (see Nomura et al. 
2007). The H$_2$ cosmic-ray ionisation rate
is set to be $\zeta_{\rm CR}=1.3\times 10^{-17}$ s$^{-1}$, and the
attenuation is neglected since it is not significant except very
close to the disk midplane at the disk radii of $x\sim 1$AU in this model.
%a H$_2$ cosmic ray ionisation rate of $1.3\times 10^{-17}$ s$^{-1}$ is
%adopted. 
%
%The chemical reactions in the gas-phase are assumed to be initiated by
%the evaporation of ice mantle molecules on dust grains which are
%transported from the cold outer disk into the hot inner disk. 
As the initial condition of the calculations,
the fractional abundances of the species are taken from
%Table 1 of 
Nomura \& Millar (2004; hereafter NM04), in which 
the initial ice mantle composition is fixed so as to be
consistent with the infrared absorption features of ices
observed towards young stellar objects as
well as the line emission of gas-phase molecules observed towards a hot
core, G34.3+0.15 (see Table 1).

%
%As we particularly focus on the observational properties of a
%methanol line in Sect. 4, we simply start the calculations where most
%methanol evaporates into the gas, that is, where the
%timescale of the thermal evaporation of methanol, $\tau_{\rm evap}$,
%becomes shorter than its adsorption time onto dust grains, $\tau_{\rm
%ads}$ 
%%in this model 
%(see also Fig.~\ref{f4}, and see e.g., 
%Nomura \& Millar 2004 and references therein
%%Tielens \& Allamandola 1987 
%for $\tau_{\rm evap}$ and $\tau_{\rm ads}$). 
%
%Now, the evaporation time is proportional to
%$\exp(E_{b}/kT_d)$ (see e.g., Tielens \& Allamandola 1987 for $\tau_{\rm
%evap}$ and $\tau_{\rm ads}$), 
%and adsorption times are given by $\tau_{\rm
%evap}=\nu_{0}^{-1}\exp(E_{b}/kT_d)$ and $\tau_{\rm ads}=(S\pi a^2
%d_gnv_{\rm th})^{-1}$, respectively, 
%%where $E_{b}$ is the
%%binding energy of the molecules onto the dust grains, $\nu_{0}$ its 
%%vibrational frequency, $kT$ the thermal energy of the gas, $S$ the
%%sticking probability, $\pi d_ga^2n$ the grain surface area per unit
%%volume, $v_{\rm th}$ the thermal velocity of the molecule 
%and the icy mantle evaporation is very sensitive to the dust temperature,
%$T_d$, and the binding energy of a molecule onto dust grains, $E_{b}$.
%(see NM04 and references therein for details). 

Since we particularly focus on the observational properties of a
methanol line in Sect.~4, we simply start the calculations where most
methanol evaporates into the gas, that is, where the
timescale of the thermal evaporation of methanol, $\tau_{\rm evap}$,
becomes shorter than its adsorption timescale onto dust grains, $\tau_{\rm
ads}$ (see Fig.~\ref{f4}). 
The evaporation time 
is given by $\tau_{\rm evap}=\nu_{0}^{-1}\exp(E_{b}/kT_d)$, 
where $\nu_{0}$ is the vibrational frequency of the adsorbed species on
the grain surface. The timescale is very
sensitive to the binding energy of the molecules onto the dust grains
$E_{b}$, and the thermal energy of the grains, $kT_d$. 
The binding energy of methanol of $E_b/k=4235$K is used in this work,
which is experimentally measured for pure methanol ice (Sandford
\& Allamandola 1993). We note that the binding energy of methanol
on methanol is equal to that of methanol on water ice within the error
bars (W.A.Brown, private communication).  
The adsorption time is given by $\tau_{\rm ads}=(S\pi a^2 d_gnv_{\rm
th})^{-1}$, where $S$ is the
sticking probability, $\pi a^2d_gn$ the grain surface area per unit
volume ($d_g$ is the abundance of the dust grains relative
to total hydrogen nuclei), and
$v_{\rm th}$ the thermal velocity of the molecule 
(e.g., Tielens \& Allamandola 1987). Here we use $S=0.3$ and
the averaged grain surface area per total hydrogen nuclei of
$<d_ga^2>=2.2\times 10^{-22}$cm$^2$ which is similar to that 
of the MRN dust model by Mathis et al. (1977) (see NM04 for details of
the parameters; see also discussions in Sect.~5 for effect of dust
evolution).  
In this model all the ice mantle molecules evaporate at the same dust
temperature of $T_d\sim 85-100$K, depending on the local density.
The influence of this simple treatment of the ice mantle evaporation
is briefly discussed in Sect.~5.
%
%and the icy mantle evaporation is very sensitive to the dust temperature,
%$T_d$, and the binding energy of a molecule onto dust grains, $E_{b}$.
%(see NM04 and references therein for details). 
%
%For simplicity, all the icy mantle molecules are assumed to evaporate at 
%the dust temperature of $T_d\sim 85-100$K
%in these calculations, but 
%The evaporation of ice mantle molecules, however, may depend on the
%binding energy of each molecule 
%and the morphology of water ice (e.g., Collings et al. 2004). 
%Therefore, different parent molecules may evaporate at different
%positions (see e.g., Markwick et al. 2002). 
%In addition, surface chemistry involving heavy radicals on warm dust
%grains may affect molecular abundances in both solid- and gas-phase
%(e.g., Garrod \& Herbst 2006; Aikawa et al. 2008).
%More detailed model including the detailed evaporation process
%as well as surface reactions on dust grains is to be constructed in
%future work for more broaden and complete analysis.

\begin{figure}
% \centering
% \includegraphics[width=8cm]{f1a.eps}
% \includegraphics[width=8cm]{f1b.eps}
\hspace*{0.8cm} \includegraphics[width=13cm]{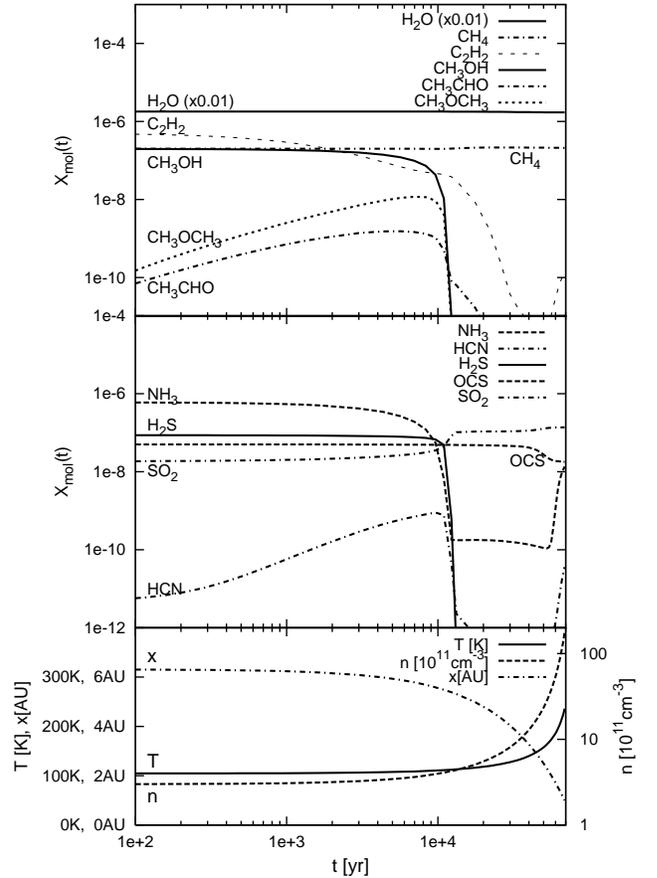}
 \caption{Evolution of molecular abundances relative to total
 hydrogen nuclei along the accretion flow, 
 $z=1.2H$, for a model with the mass accretion rate of
 $\dot{M}=1.0\times 10^{-8}M_{\odot}$ yr$^{-1}$. 
The figure at the bottom shows the gas
 density ($n$) and temperature ($T$) profiles and the distance from the
 central star ($x$). The molecules
 evaporated from ice mantles are destroyed by chemical reactions around
 the timescale of $\sim 10^4$yr.}
 \label{f1}
\end{figure}

\section{Resulting molecular distributions}

We calculate molecular abundances
using the obtained density and temperature profiles.
%Making use of the density and temperature profiles of the disk, we
%calculate the chemical reactions along the accretion flows, which are
%initiated by the evaporation of icy mantle molecules on dust grains.
Figure~\ref{f1} shows the resulting evolution of abundances of some
molecules relative to total hydrogen nuclei along the stream line
of $z=1.2H$. The results for 
a fiducial model with the mass accretion 
rate of $\dot{M}=1.0\times 10^{-8}M_{\odot}$ yr$^{-1}$ are plotted,
in which the gas density and temperature range over $n\sim 3\times
10^{11}-1\times 10^{13}$ cm$^{-3}$ and $T\sim 100-240$K, respectively, 
along the flow. The gas density ($n$) and temperature ($T$)
profiles as well as the distance from the central star ($x$) are also
plotted in the figure at the bottom. 

Figure~\ref{f1} shows that the parent molecules which evaporate from
grain surface, such as NH$_3$, H$_2$S, C$_2$H$_2$ and
CH$_3$OH in this model, are relatively stable, but are eventually
destroyed by reactions with ionised 
species or atomic hydrogen to produce daughter species, such as HCN,
SO$_2$, and CH$_3$OCH$_3$, around a timescale of $\tau_{\rm chem}\sim
10^4$yr. 
The parent and a part of daughter species end up as CO and small
hydro-carbon molecules.
The timescales for the destruction of these parent molecules are nearly
independent of the total density, $n$.
This is because the number densities of ionised species and atomic
hydrogen, which are formed via cosmic-ray ionisation and induced
photoreactions, respectively, are not very sensitive to $n$. 
OCS is destroyed on a longer timescale than other parent molecules
since its main 
reactant is H$_3^+$, while many of the other molecules are destroyed
mainly by H$_3$O$^+$ which is more abundant than H$_3^+$. OCS does not
react with H$_3$O$^+$ because its proton affinity is smaller than that
of water. 
%In addition, the formation rates of OCS and CH$_4$ are relatively high
%compared with the other molecules.
%
H$_2$O and CH$_4$ are hardly altered since they are produced efficiently
in the gas-phase, especially at high temperature in the case of H$_2$O,
which is formed via endothermic reactions, O $\stck{{\rm H}_2}$ OH
$\stck{{\rm H}_2}$ H$_2$O. Methane is formed via the destruction of
methanol and the formation processes are less temperature dependent.
The abundances of NH$_3$ and HCN increase around $t\sim 10^5$yr in
Figure~\ref{f1} owing to endothermic reactions and high temperature in
the inner disk.
The detailed gas-phase chemistry induced by the evaporated
molecules is similar to that occurring in hot molecular cores, which is
discussed in NM04. 
Along $z=1.2H$ molecules are not affected by UV and X-ray irradiation
from the central star in this model (see also Sect.~\ref{S4}).
We note that the timescales for destruction of these parent molecules by
chemical reactions have uncertainties of about an order of
magnitude due to the uncertainties in the rate coefficients of the
reactions (e.g., Wakalem et al. 2005). 

\begin{figure*}
 \centering
 \includegraphics[width=8cm]{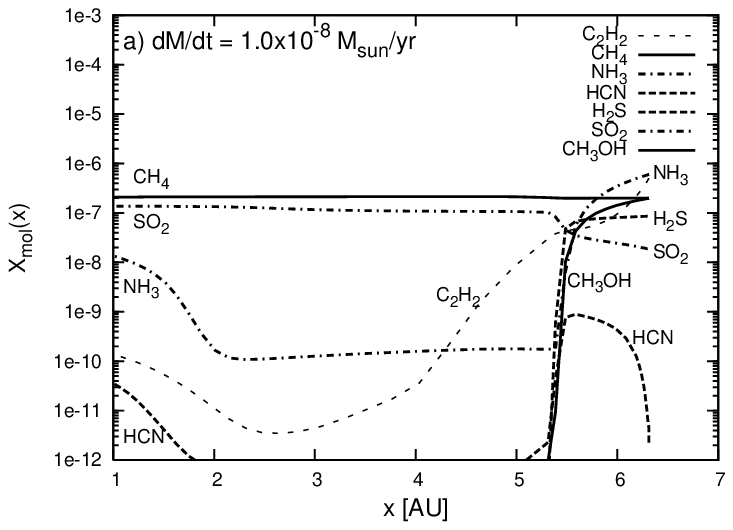}
 \includegraphics[width=8cm]{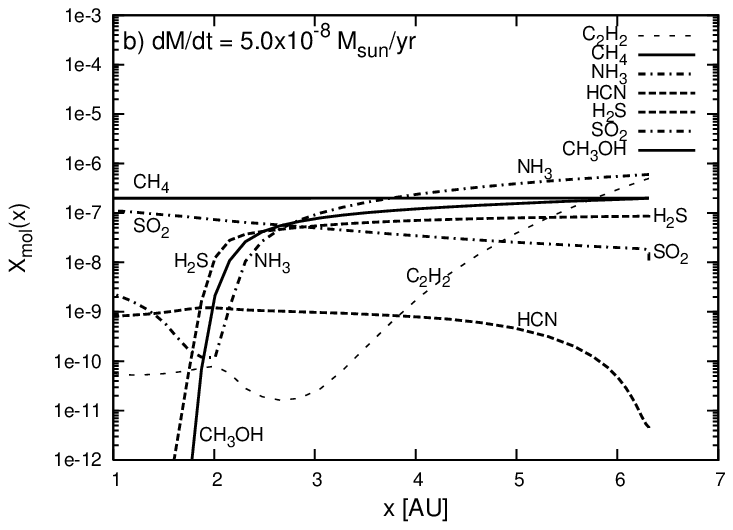}
 \caption{Abundance profiles of some species relative to total
 hydrogen nuclei as a
function of the disk radius along the stream line of $z=1.2H$ for models
with the mass accretion rates of $\dot{M}=$ (a) $1.0\times 10^{-8}$
and (b) $5.0\times 10^{-8}M_{\odot}$ yr$^{-1}$.
Some molecules are abundant just around the region of the ice mantle
 evaporation when $\dot{M}$ is low and
the accretion time ($\tau_{\rm acc}$) is longer than the timescale of
 the chemical reactions which destroy the parent molecules,
%velocity, $v_{\rm acc}$, is low, 
while they are abundant even close to the central star when 
$\dot{M}$ is high and $\tau_{\rm acc}$ is short enough.}
%$v_{\rm acc}$ is high.}
 \label{f2}
\end{figure*}

In Figure~\ref{f2}, we plot the abundance profiles of some species 
relative to total hydrogen nuclei as a 
function of the disk radius along the stream line of $z=1.2H$ for models
with mass accretion rates of $\dot{M}=$ (a) $1.0\times 10^{-8}$
and (b) $5.0\times 10^{-8}M_{\odot}$ yr$^{-1}$.
In the steady accretion flow, a location of a fluid particle in the
disk, $x$, is simply related to a time after the ice evaporation, $t$,
as $t=\int_{x_{\rm evap}}^x \{v_{\rm acc}(x')\}^{-1}dx'$ (where $x_{\rm
evap}$ is the evaporation radius), as shown in the bottom of
Figure~\ref{f1}. 
The radial dependence of the accretion velocity is weak, and
$v_{\rm acc}\sim 40$ cm s$^{-1}$
for the fiducial model with $\dot{M}=1.0\times 10^{-8}M_{\odot}$
yr$^{-1}$. In this work the accretion velocity is set to be proportional
to the mass accretion rate ($v_{\rm acc}\propto \dot{M}$)
by artificially fixing the surface density of
the disk to that of the fiducial model for the purpose of clear
understanding of the effect of accretion flow on the chemical structure
in the disk. 
%the surface density of the disk is artificially fixed with
%that of the fiducial model so that the accretion velocity
%is proportional to the mass accretion rate for clear understanding of
%the effect of the accretion flow on the chemical structure in the disk.

%__________________________________________________ Table 2
%\begin{table}[b]
%\caption[]{Observed fractional abundances of warm molecules with respect
% to CO.}\label{T2}
%$$ \begin{array}{p{0.11\linewidth}|ccc}
%%{p{0.11\linewidth}cp{0.12\linewidth}c} 
%\hline 
% Species & {\rm IRS\ 46} & {\rm GV\ Tau\ S} & {\rm AA\ Tau} \\ \hline
% H$_2$O & - & - & 1.3 \\
% OH & - & - & 0.18 \\
% HCN & 0.025 & 0.0031 & 0.13 \\
% C$_2$H$_2$ & 0.015 & 0.0062 & 0.016 \\
% CO$_2$ & 0.05 & - & 0.004-0.26 \\
% CH$_4$ & - & < 0.0019 & - \\ \hline
%\end{array}
%$$ 
%\end{table}
%

Figure~\ref{f2} shows that the molecular abundance profiles in the inner
disk change 
dramatically depending on the velocity of the accretion flow. When the 
accretion velocity is small and the accretion time, $\tau_{\rm acc}\sim
x/v_{\rm acc}$, is longer than the timescale of the chemical reactions
which destroy the evaporated molecules, $\tau_{\rm chem}$, the
parent molecules are transported only a small distance along the
accretion flow before they are destroyed by the chemical reactions. 
Therefore, some parent and daughter species are abundant only in the
region where parent molecules are evaporated into the gas. 
On the other hand, if the accretion velocity is high and the accretion 
time is shorter than the chemical timescale, the parent 
molecules are transported to the central star before they are destroyed,
and high abundances occur even close to the star.
Some daughter species, such as HCN, become abundant gradually 
as the gas is transported inwards
%only in the very inner region of the disk 
in this case. We note that the abundances of some daughter species
%, such as HCN and SO$_2$, 
are very sensitive to the initial fractional abundances;
%some initial
%conditions, such as the ice mantle composition and the H$_2$O
%abundance in the gas-phase; 
for example, if atomic nitrogen is abundant
initially, the maximum value of the HCN abundance increases by about two
orders of magnitude compared with the model we used here.
%and large differences appear in their abundances for different conditions. 
%
Along the stream line of $z=1.2H$, C$_2$H$_2$ is destroyed by an
endothermic reaction, C$_2$H$_2$+O$\rightarrow$CO+CH$_2$. 
%which is sensitive to the gas temperature. 
Thus, the C$_2$H$_2$ abundance decreases, independent of the accretion
rate, around $x\sim 4-5$AU, where the gas temperature becomes high
enough for the above reaction to proceed. 
%and drops more quickly than other parent molecules for the model
%with $\dot{M}=5.0\times 10^{-8}M_{\odot}$ yr$^{-1}$. 
%by a reaction with atomic oxygen 
%where the density and temperature are relatively high.
Along stream lines in upper layers of the disk slightly above $z=1.2H$, 
where the gas density is relatively low, C$_2$H$_2$ is destroyed by
reactions with molecular ions around a timescale of $10^4$yr, similar to
other parent molecules. 
\section{Methanol line emission}\label{S4}

\begin{figure}
 \centering
 \includegraphics[width=9cm]{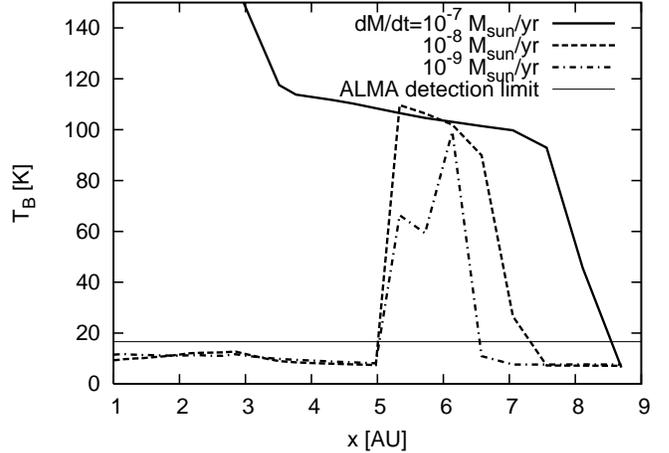}
 \caption{Brightness temperature of the 145GHz $J_K=3_0-2_0$ line of
 methanol calculated along a single line of sight as a function of
 the disk radius for models with mass 
accretion rates of $\dot{M}=1.0\times 10^{-7}$ ({\it solid line}),
 $1.0\times 10^{-8}$ ({\it dotted line}), and $1.0\times 10^{-9}$ ({\it
 dot-dashed line}) $M_{\odot}$ yr$^{-1}$. The ALMA detection limit for
 the beam size of $\sim 14$AU is 
 also plotted with a thin solid line, showing that the line emission
 will be detectable in the inner disk when the accretion rate and
 the accretion velocity are high.}
 \label{f3}
\end{figure}

Using the gas density and temperature as well as the molecular abundance
profiles obtained in Sect.~3, we calculate the brightness temperature of
145GHz $J_K=3_0-2_0$ transition of methanol. 
%which is an abundant ice molecule in many  
%%Now, methanol has been observed in ice absorption features towards many
%high and low mass star forming cores (e.g., Gibb et al. 2004; Boogert et
%al. 2008). 
%Methanol, which is difficult to form efficiently in the gas phase, 
%is also known to be more
%%Observations of molecular line emission, on the other hand, have shown
%%that methanol is 
%abundant in the gas-phase around high and low mass
%young stars (e.g., Macdonald et al. 1996; Sch\"oier et al. 2002) 
%%while it is less abundant 
%by a few orders of magnitude compared with cold dark clouds 
%%in contrast 
%(e.g., Ohishi et al. 1992). 
%In addition, laboratory experiments have shown that methanol can 
%be formed in ice mantle with reasonably short timescale 
%as a result of hydrogenation of carbon monoxide 
%%and the observed abundances of icy methanol can be
%%reproduced by the reaction with a reasonable timescale 
%(e.g., Watanabe et al. 2006). 
%Therefore, methanol is thought one of the most probable components of
%grain mantle molecules.
%
Figure~\ref{f3} shows the resulting brightness temperature of
%$T_B$, of our calculations 
%at the line centre of the 145GHz $J_K=3_0-2_0$ transition
%of methanol 
the line along a single line of sight
as a function of the disk radius for models with the
mass accretion rates of $\dot{M}=1.0\times 10^{-7}$
({\it solid line}), $1.0\times 10^{-8}$ ({\it dotted
line}), and $1.0\times 10^{-9}$ ({\it dot-dashed line}) $M_{\odot}$
yr$^{-1}$. We assume that
%The molecules are simply assumed to be 
the molecules are in the local thermal equilibrium.
The assumption is reasonable since
the line is emitted from regions 
%in the disk 
where the density is
much higher than the critical density for this transition, $n_{\rm cr}\sim
2\times 10^5$cm$^{-3}$ (e.g., Pavlychenkov et al. 2007).
The disk is assumed to be face-on to an observer.
%and the disk is faced-on to an observer.
The molecular data is taken from Leiden Atomic and Molecular Database
(http://www.strw.leidenuniv.nl/\textasciitilde moldata/)
(Sch\"oier et al. 2005), and a part of the RATRAN code 
(http://www.sron.rug.nl/\textasciitilde vdtak/ratran/ratran.html)
(Hogerheijde \& van
der Tak 2000) is used for the line radiative transfer calculations (see
NM05 for the detailed equations used in the calculations).

The figure shows that the brightness temperature of the line, $T_B$, is
high all over the inner disk when the accretion velocity is high, while
$T_B$ becomes high just around the region where the methanol is
evaporated into the gas when the accretion velocity is low, since $T_B$
traces the abundance of methanol (see Sect.~3).
Now, methanol is mainly destroyed by chemical reactions with H$_3$O$^+$, 
%where the density is high,
on a timescale around $\tau_{\rm chem}\sim (kn_{{\rm H}_3{\rm
O}^+})^{-1}\sim 
10^4$yr. Here the rate coefficient is $k\sim 3\times 10^{-9}$cm$^3$s$^{-1}$
and the number density of H$_3$O$^+$ is $n_{{\rm H}_3{\rm O}^+}\sim
10^{-3}$ cm$^{-3}$,
%$n_{{\rm H}_3^+}\sim 10^{-4}$ cm$^{-3}$. 
which is almost independent of the total number density, $n$,
since the formation process of H$_3$O$^+$ is related to cosmic-ray
ionisation. 
The timescale of methanol destruction is slightly affected by the
abundance of NH$_3$ which helps recycle methanol by
reacting with CH$_3$OH$_2^+$ (e.g., Rodgers \& Charnley 2001). 
When the accretion rate is high, $T_B$ is high even in slightly outer
regions ($x\sim$7--9 AU)
because the dust temperature is higher due to the viscous heating.

\begin{figure}
 \centering
 \includegraphics[width=9cm]{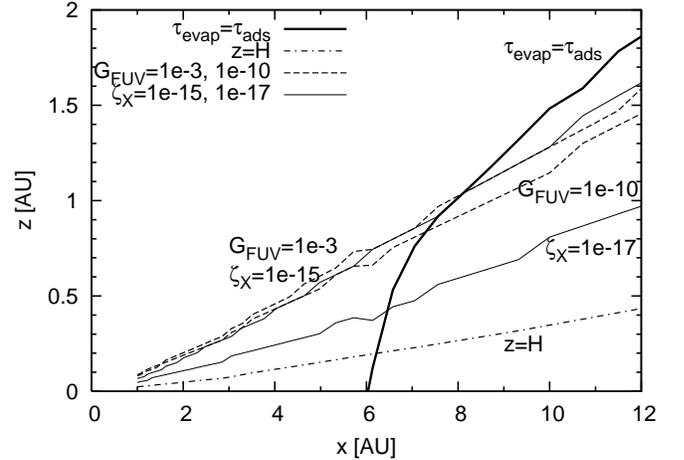}
 \caption{Line of $\tau_{\rm evap}=\tau_{\rm ads}$ inside which
 the icy methanol evaporates into the gas ({\it thick solid
 line}), line of $z=H$ ({\it dot-dashed line}), and contour lines of
 $G_{\rm FUV}=10^{-3}$ and $10^{-10}$ ({\it thin solid lines}),
 $\zeta_{\rm X}=10^{-15}$ and $10^{-17}$ s$^{-1}$ ({\it dashed
 lines}) for the fiducial model with $\dot{M}=1.0\times 10^{-8}
 M_{\odot}$ yr$^{-1}$. The evaporated methanol is quickly
 dissociated due to strong UV photons or X-rays at $x\geq 7$AU.}
 \label{f4}
\end{figure}

At the disk surface methanol is photodissociated due to UV and
X-ray irradiation from the central star. The timescales of the
dissociation are roughly given by $\tau_{\rm UV}\sim k_{\rm UV}^{-1}\sim
30G_{\rm FUV}^{-1}$ yr and $\tau_{\rm X}\sim k_{\rm X}^{-1}\sim
10^{-11}\zeta_{\rm X}^{-1}$ yr, respectively, 
%Here $k_{\rm UV}\sim 10^{-9}G_{\rm FUV}$ s$^{-1}$ and $k_{\rm X}\sim
%3\times 10^3\zeta_{\rm X}$, 
where $G_{\rm FUV}$ is the FUV radiation field normalized by 1.6$\times
10^{-3}$ ergs s$^{-1}$ cm$^{-2}$ and $\zeta_{\rm X}$ is the total
hydrogen ionisation rate by X-rays.
Therefore, the photodissociation dominates the destruction of methanol
where 
%the FUV radiation field is 
$G_{\rm FUV}\ga 3\times 10^{-3}$
%[$G_{\rm FUV}$ is normalized by 1.6$\times 10^{-3}$ ergs s$^{-1}$
%cm$^{-2}$] and the X-ray ionisation rate is 
or $\zeta_{\rm X}\ga 10^{-15}$ s$^{-1}$. 
In Figure~\ref{f4} the contour lines of $G_{\rm FUV}=10^{-3}$ and
$10^{-10}$ ({\it thin solid lines}), $\zeta_{\rm X}=10^{-15}$ 
and $10^{-17}$ s$^{-1}$ ({\it dashed lines}) are plotted in $z$ vs $x$
plane of the disk together with the line of $\tau_{\rm evap}=\tau_{\rm
ads}$ for methanol ({\it thick solid line}), which corresponds to $T_d\sim
85-100$K, and $z=H$ ({\it dot-dashed line}) for the fiducial model. 
Figure~\ref{f4} shows that 
the line brightness temperature at $x\ga 6-7.5$AU in Figure~\ref{f3}
is low because the methanol can evaporate from dust grains only in the
surface layer of the disk where it is dissociated
quickly by strong X-ray and UV irradiation from the central star.
%In this region $T_B$ becomes high only when
%%when the accretion velocity is high enough and 
%$\tau_{\rm acc}$ is shorter than $\tau_{\rm X}$ and $\tau_{\rm UV}$.

%
%The turbulent mixing in the radial direction may not be very significant
%will transport material
%inward and outward only if the material gain or lose angular momentum.
%if there is no effective angular momentum transfer process.
%Detailed models which takes into account angular
%momentum transfer of the material should be constructed in future work. 

In Figure~\ref{f3} we also plot with a thin solid line the ALMA detection limit
for the methanol line, calculated
using the ESO ALMA Sensitivity Calculator 
(http://www.eso.org/sci/facilities/alma/observing/tools
/etc/index.html).
The detection limit of $\sim 17$K is achieved when we observe for 600 sec
by using 50 antennas of the 12m Array, assuming a velocity resolution of
1 km s$^{-1}$, and a beam size of 0.1 arcsec, which corresponds to
$\sim$14AU when the disk is located at the distance of nearby
star-forming regions, $\sim$ 140 pc. 
%
%{\bf We note that the strong line emitting region (where $T_B>100$K) is
%larger than $\sim$14AU
%as shown in Figure~\ref{f3}}
%
%An unsuccessful search for the 145GHz $J_K=3_0-2_0$ transition of
%methanol has been made 
The 145GHz $J_K=3_0-2_0$ transition of methanol has not been detected
towards DM Tau (Dutrey 2001), and other methanol line searches 
towards some T Tauri and Herbig Ae stars (Thi et al. 2004; Semenov
et al. 2005), have also been unsuccessful.
%but no detection has been reported yet. 
%
Our results suggest that when the accretion velocity at $x\leq 10$AU 
%in the disk 
is high enough, ALMA can detect the methanol line, taking advantage of
high spatial resolution to avoid beam dilution.
%observations with high spatial resolution and high sensitivity will
%detect the methanol line without suffering from beam dilution.
%(where the disk is located at the distance of $\sim$ 140 pc).
When the accretion velocity is low enough, line detection will be
difficult even with ALMA. Thus, the
observations of methanol by ALMA
could be a useful tool to diagnose the existence of accretion flows 
in the inner disks. 
%in the inner region of the disks. 
%
%We note that the line will no trace the accretion flow  at the disk
%midplane since it is optically thick in this region. Less abundant
%isotopes could trace the flow at the disk midplane.

%uncertainties in disk structure (surface density)
%-> uncertainties in accretion rates

\section{Discussions}

In this section, we discuss some possible effects of the
assumptions on our results.
%which we have made in this work.

First, we discuss the influences of the simple treatment of
the evaporation process we have taken in this work.
Although we simply assume that all the species evaporate at the same
dust temperature, the evaporation of ice mantle molecules may depend on
the binding energy of each molecule 
and the morphology of water ice (e.g., Collings et al. 2004). 
Therefore, different parent molecules may evaporate at different
positions (see e.g., Markwick et al. 2002), which will affect the
spatial distributions of molecular abundances.
In particular, CO and N$_2$ are likely to evaporate at temperatures much
lower than 85K, particularly if they are abundant in the surface layers 
of the ice. However, given the chemical stability of these species, they 
are unlikely to affect the overall abundance of a molecule such as methanol.
%a species to which we return in the following section.
So, the timescales for the destruction of the parent molecules by chemical
reactions, $\tau_{\rm chem}$, analysed in Sect.~3
%above 
%on the other hand, 
will not be very sensitive to the treatment of the ice mantle evaporation.
In addition to the evaporation process, surface chemistry involving
heavy radicals on warm dust 
grains may affect molecular abundances in both solid- and gas-phase
(e.g., Garrod \& Herbst 2006; Aikawa et al. 2008).
More detailed model including the detailed evaporation process
as well as surface reactions on dust grains should be constructed in
future work.

Next, if we take into account the effect of turbulent mixing in the vertical
direction, the methanol abundance may increase near the line, $\tau_{\rm
evap}=\tau_{\rm ads}$, (see Fig.~\ref{f4}) when the diffusion time is
shorter than the timescale of the photodissociation (e.g., Willacy et
al. 2006). In the inner disk ($x\la 6$AU) the methanol abundance may
slightly decrease via the vertical mixing because of the
photodissociation at the disk surface. 
The turbulent mixing in the radial direction may enhance the gas-phase
methanol abundance outside the evaporation radius if the timescale of
the turbulent diffusion is shorter than the timescale of the chemical
reactions and that of the adsorption on dust grains. A detailed
mixing model which treats angular momentum 
transfer will be needed in order to obtain the diffusion timescale, and
advance further discussion. 

Finally,
we note that the brightness temperature of the methanol line will 
reflect some conditions other than the accretion velocity: the
high-$T_B$ region becomes smaller if the central star is less luminous
and the dust temperature in the disk is low. Also, $T_B$ becomes low if
the methanol abundance in the ice mantle is low. 
%{\bf or the binding energy of the methanol is high}. 
If the amount of small
dust grains decreases due to the dust coagulation and settling in the
disk, methanol is photodissociated by
ineffectively attenuated FUV photons from the central star
(e.g., Aikawa \& Nomura 2006), and therefore $T_B$ drops,
although the evaporation radius moves slightly outwards owing to
the decrease of the total grain surface area per unit volume, that is,
the decrease of the adsorption rate (e.g., Aikawa 2007). 
%also in this case. 
Thus, some causes other than low 
accretion velocity can lead to low methanol abundance and the
non-detection of the line. However, if the observed methanol abundance
is high in the 
inner disk, then the accretion velocity must be high in the planet-forming
region.

\section{Conclusions}

We have calculated the gas-phase chemical reactions, initiated by the ice mantle
evaporation, along the accretion
flow in protoplanetary disks to show that the accretion velocity affects the abundance
profiles of the evaporated molecules and their daughter species in the
inner disk. 
%depending on whether the accretion time is shorter or
%longer than the timescale of destruction of the evaporated molecules 
%by chemical reactions.
%
We have obtained the brightness temperature profile of an
emission line of methanol, and suggested that this could be useful
as an indicator of accretion velocity in the planet-forming regions
in disks using ALMA observations.
% with high sensitivity and
%high spatial resolution. 

\begin{acknowledgements}
We are thankful to an anonymous referee and an editor, M. Walmsley, for
 their comments which improved the clarity of our discussion.
H.\ N. is financially supported by the JSPS Postdoctoral Fellowships
 for Research Abroad. Astrophysics at QUB is supported by a grant from the STFC.
\end{acknowledgements}

\end{document}